\documentclass[aps,prl,toolkits,twocolumn]{revtex4}

\usepackage{graphicx,bm}
\usepackage{amsmath,amssymb}
\usepackage{mathrsfs,verbatim,subfigure}
\usepackage{graphics}

\renewcommand\section[1]{{\em #1 ---}}

\begin{document}

\voffset=1cm

\title{Realistic Implementation of Massive Yang Mills for $\rho$ and $a_1$ Mesons}

\author{Paul M. Hohler}
\email{pmhohler@comp.tamu.edu}
\affiliation{Cyclotron Institute and Department of Physics\,\&\,Astronomy,
Texas A{\&}M University, College Station, Texas 77843-3366, USA }

\author{Ralf Rapp}
\email{rapp@comp.tamu.edu}
\affiliation{Cyclotron Institute and Department of Physics\,\&\,Astronomy,
Texas A{\&}M University, College Station, Texas 77843-3366, USA }

\date{\today}

\begin{abstract}
We revisit the massive Yang-Mills approach to implement axial-/vector mesons into
the chiral pion Lagrangian. Employing the non-linear realization of chiral symmetry,
we compute vector and axial-vector spectral functions. This includes, for the first
time, a resummation of the $\rho$ propagator in the $a_1$ selfenergy while maintaining
a partially conserved axial-vector current (PCAC). A realistic $\rho$ propagator,
combined with a recent idea of a chirally invariant continuum in the vector and axial-vector
channels, turns out be critical to obtain a quantitative description of hadronic
$\tau$-decay data which has not been achieved before in local-gauge approaches to
the chiral lagrangian. The thus obtained model provides a realistic basis to
rigorously address the long-standing question of chiral symmetry restoration in
the context of dilepton data in heavy-ion collisions.
\end{abstract}

\pacs{}
\maketitle

Chiral effective theories are a successful and widely used tool to describe the
low-energy limit of Quantum Chromodynamics (QCD) dominated by
pions~\cite{Gasser:1983yg}. Several approaches have been pursued to extend the
chiral Lagrangian by including the light vector (V) and axial-vector (AV) mesons,
$\rho$ and $a_1$. A theoretically appealing framework is their implementation as local
gauge bosons of the chiral group, $SU(2)_L \times SU(2)_R$, with explicit mass
terms~\cite{Meissner:1987ge}. The universality of the gauge principle
limits the free parameters while recovering the phenomenologically
successful concept of vector meson dominance (VMD). Two leading realizations of this
idea are the massive Yang-Mills (MYM)~\cite{Gomm:1984at} and hidden local symmetry
(HLS)~\cite{Bando:1987br,Harada:2003jx} schemes which turn out to be on-shell
equivalent.

Early studies using tree-level MYM were successful in reproducing basic decay
properties of $\rho$ and $a_1$ mesons (once so-called non-minimal terms were
included)~\cite{Gomm:1984at,Song:1993ae,Ko:1994en}, but recent precision data for
axial-/vector spectral functions from hadronic $\tau$
decays~\cite{Barate:1998uf,Ackerstaff:1998yj} pose more stringent constraints,
and require loop-level calculations. These lead to several problems for MYM; {\it e.g.},
in the AV channel, the $a_1 \rightarrow \rho\pi$ width develops a zero in the
experimentally accessible region, followed by a rapid growth at higher energies.
Furthermore, the basic chiral coupling of the $a_1$ to external electro-weak (EW)
fields entails a lack of strength in the AV spectral function. These issues have
hampered attempts to describe the $\tau$-decay data. Within HLS, extensive 1-loop
calculations have been carried out, showing good agreement with many decay
branchings~\cite{Harada:2003jx}, but a comparison to the experimental spectral
functions is not available.

Recent investigations have therefore either given up on the local implementation
of the gauge group in favor of a global one~\cite{Urban:2001ru,Parganlija:2010fz},
or abandoned the notion of $\rho$ and $a_1$ as gauge bosons~\cite{Wagner:2008gz},
thus decoupling their role as chiral partners.
A global implementation removes constraints between couplings of different terms
in the effective Lagrangian. The added freedom allows for a successful fit of the
AV spectral function to $\tau$ data while preserving chiral symmetry at the 1-loop
level~\cite{Urban:2001ru}, or to achieve a good description of a wide range of
light-meson properties at tree level~\cite{Parganlija:2010fz}.  In
Ref.~\cite{Wagner:2008gz}, the $a_1$ is dynamically generated from $\rho\pi$
interactions which also allows for a good fit of the AV $\tau$-decay data.
However, as we will show below, it is also possible to resolve the problems within MYM.

In this letter, we implement two novel aspects
into MYM: a fully dressed $\rho$ propagator in the $a_1$ selfenergy and a chirally
invariant high-energy continuum. Preserving chiral symmetry is challenging within
the dressing procedure. We develop criteria by which particular
vertex correction diagrams can be identified and included so as to preserve the chiral
properties. In practice, the broad $\rho$ smears out the energy dependence of
the $a_1$ width, thus eliminating the troublesome zero. A universal perturbative
continuum was recently suggested in a phenomenological application of V and AV
spectral functions to QCD and Weinberg sum rules~\cite{Hohler:2012xd}, and was
found to extend into the $a_1$ peak region. This extra strength further helps in
arriving at a quantitative description of the measured spectral functions within MYM.

The MYM Lagrangian with non-minimal terms reads
\begin{equation}
\begin{split}
\mathcal{L}_{\rm MYM}& = \frac{\tilde{f}_\pi^2}{8} ( {\rm Tr}[D_\mu U^\dag D^{\mu} U]
+ \tilde{m}_\pi^2{\rm Tr}[U+U^\dag-2] )\\
& - \frac{1}{2} {\rm Tr}[F_L^2 + F_R^2]  + m_0^2 {\rm Tr}[A_L^2+A_R^2] \\
& - i \xi \ {\rm Tr}[D_\mu U D_\nu U^\dag F_L + D_\mu U^\dag D_\nu U F_R]\\
& +\gamma \ {\rm Tr}[F_L U F_R U^\dag] \ ,
\end{split}
\end{equation}
where the pion fields $\pi = \pi^a \tau^a/\sqrt{2}$ emerge from expanding the nonlinear
realization of the field $U$=$\exp\left [{2i}/{F_\pi} \pi \right ]$.
The covariant derivative is defined via the gauge coupling, $g$, and the gauge
bosons, $A_{L/R} = A_{L/R}^a \tau^a /\sqrt{2}$, as
\begin{equation}
D_\mu U= \partial_\mu U - i g (A_{L \mu} U - U A_{R \mu}) \ .
\end{equation}
The $\rho$ and $a_1$ fields are identified as the vector and axial-vector
combinations of the gauge bosons according to
\begin{equation}
\begin{split}
\rho_\mu &= \kappa_v(A_{L \mu}+A_{R \mu}) \
\\
a_\mu &=\kappa_a(A_{L \mu} - A_{R \mu}) \ ,
\end{split}
\end{equation}
where $\kappa_{v/a} = \left(1 \mp \gamma \right)^{-1/2}$ are field renormalizations
needed to recover the kinetic terms of the meson fields. The chiral and gauge
symmetries are explicitly broken by mass terms for the physical fields. The
shift $a_\mu \rightarrow a_\mu + \alpha Z_\pi \partial_\mu \pi$
removes the unphysical $a_\mu \partial_\mu \pi$ coupling. The pion
field renormalization, $Z_\pi$, recovers the pion kinetic term.
This also renormalizes the pion decay constant, $\tilde{f_\pi}$=$F_\pi Z_\pi$, and
the pion mass, $\tilde{m_\pi} = M_\pi/Z_\pi$, with the physical values
$F_\pi$=131\,MeV and $M_\pi$=139.6\,MeV.

The theory is coupled to external EW fields by gauging them under
the same groups as the mesons. The EW gauge couplings are fixed by the pion
charge. This leaves four parameters in the theory: the hadronic gauge coupling
$g$, the bare mass $m_0$, and the two non-minimal couplings $\gamma$ and $\xi$. They
can be traded for the phenomenologically more pertinent set of the single- and
three-derivative $\rho \pi \pi$ couplings, $g_{\rho\pi\pi}$ and
$g_{\rho\pi\pi}^{(3)}$, and bare $\rho$ and $a_1$ masses, $M_\rho$ and $M_{a_1}$.
The Lagrangian is expanded to generate the vertices of the physical fields in terms
of these four parameters.

We define the axial-/vector spectral functions as
$\rho_{V,A} = -{\rm Im} \Pi^T_{V,A}/\pi$ in terms of the 4-D transverse
current-current correlators $\Pi^T_{V,A}$. Since the mesons
and external fields are gauge bosons of the same group, the imaginary part of the
correlator can be expressed in terms of the propagator $D_{V,A}^T$ as
\begin{equation} \label{eq:vmdcor}
\rho_{V,A} = - C_{V,A} {\rm Im} D^T_{V,A} / \pi,
\end{equation}
with axial-/vector meson dominance-like couplings
\begin{equation}
C_V = \frac{M_\rho^4}{g_{\rho\pi\pi}^2}\ , \quad C_A = \frac{M_a^2 M_\rho^2}{g_{\rho\pi\pi}^2 Z_\pi^2}.
\end{equation}
Note that (A)VMD is not {\it a priori} assumed in the coupling of the EW fields, but
is recovered for the spectral function due to gauge symmetry.

A simple estimate with the above relations reveals the problem of lacking strength
in the AV channel. From the $\tau$-decay data ({\it cf.}~Fig.~\ref{fig:sf} below),
the ratio of peak heights between V and AV spectral functions is
$\rho_A(M_{a_1}^2)/\rho_V(M_{\rho}^2)$$\sim$1/3. However, evaluating Eq.~(\ref{eq:vmdcor})
with physical widths $\Gamma_\rho$$\simeq$150\,MeV, $\Gamma_{a_1}$$\simeq$450\,MeV and masses
$M_{a_1}^2\simeq Z_\pi^2 M_\rho^2 \simeq 2 M_\rho^2$, yields this ratio by a factor of
$\sim$$\sqrt{2}$ too low (more for larger $a_1$ width). In
global models, where the external fields and the axial/-vector mesons are not gauge
bosons of the same group,  the spectral functions are not constrained as in
Eq.~(\ref{eq:vmdcor}).

\begin{figure}[!t]
  \centering
	\includegraphics[width=.32\textwidth]{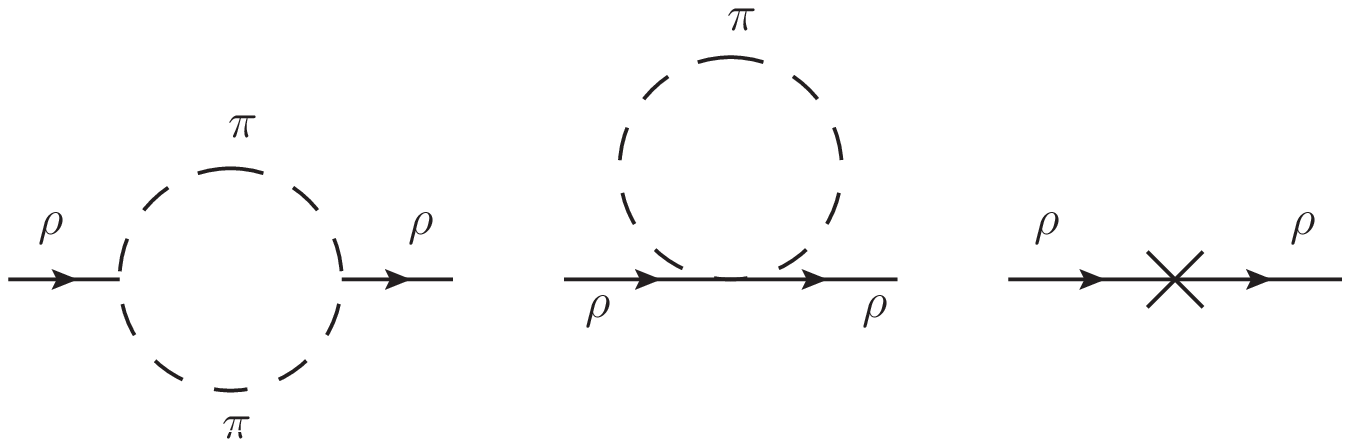}
	\includegraphics[width=.32\textwidth]{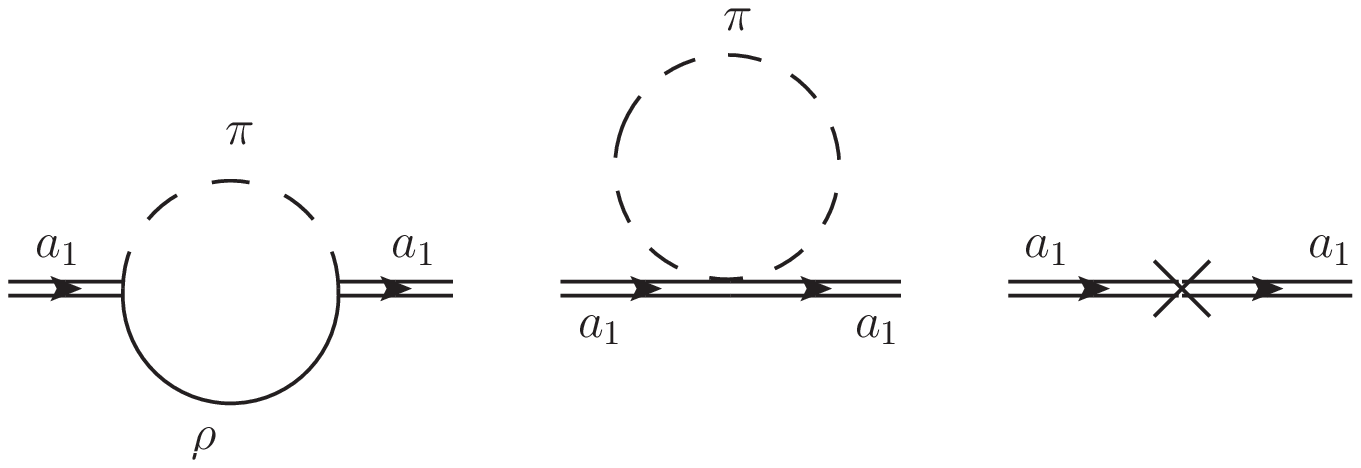}
\caption{Feynman diagrams used to calculate the $\rho$ (upper panel) and $a_1$
(lower panel) selfenergies (crosses denote counter terms).}
\label{fig:dia1}
\end{figure}

We start out by calculating the V and AV selfenergies figuring
into the propagators to 1-loop order including all diagrams as shown in
Fig.~\ref{fig:dia1}. We do not consider the $\pi a_1$ loop in the vector channel
as its threshold is well above the $\rho$ mass. This requires that, in order to
preserve gauge symmetry, only part of the $\pi$-tadpole in the vector
channel be included. The divergent loops are regulated using dimensional regulation
and suitable counter terms based on the Lagrangian, thus
preserving chiral symmetry. Some of the counter terms are fixed to renormalize
$M_\pi$ and $F_\pi$ to their physical values while others aid in fitting the
spectral functions to experimental data, corresponding to six additional free
parameters.

\begin{figure}[!b]
\centering
\includegraphics[width=.45\textwidth]{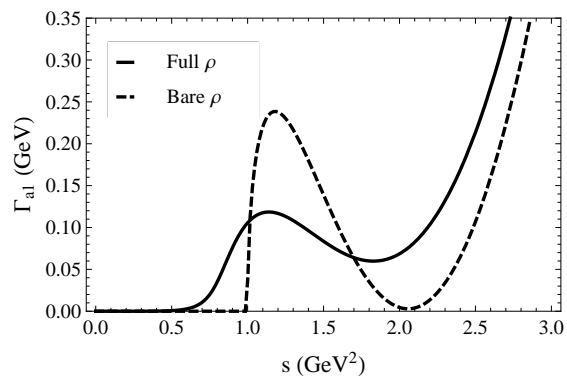}
\vspace{-0.3cm}
\caption{Energy dependence of $\Gamma(a_1 \rightarrow \rho\pi)$ with a fixed-mass $\rho$ (dashed
curve) and a resummed $\rho$ propagator with vertex corrections (solid curve).}
\label{fig:axialwidth}
\end{figure}

The resulting energy dependence of the $a_1 \rightarrow \rho\pi$ width,
$\Gamma_{a_1 \rightarrow \rho\pi}(s) = -{\rm Im} \Sigma_{a_1}^T(s)/\sqrt{s}$,
reveals another problem, as a zero at an energy near the physical $a_1$ mass
develops, {\it cf.}~the dashed line in Fig.~\ref{fig:axialwidth}. To alleviate
this issue, we will in the following incorporate the fully dressed $\rho$
propagator into the AV loop calculation. This is expected to smear out
the energy dependence and thus fill in the valley created by the zero. However,
a na\"{i}ve resummation will not preserve chiral symmetry, requiring hitherto
unknown vertex corrections.

Let us investigate the partial conservation of the axial current (PCAC).
At tree level, the axial Noether current is given in terms of the $W$, $a_1$, and
$\pi$ fields as
\begin{equation}
J_A^\mu = \alpha W^\mu+ \beta a^\mu + \gamma \partial^\mu \pi \ ,
\end{equation}
where the couplings to the axial current are given by
\begin{equation} \label{eq:nctree}
 \alpha = \frac{1}{2}F_\pi^2 Z_\pi^2 \ ,  \ \beta = \frac{1}{2}g_{\rho\pi\pi}
 F_\pi^2 Z_\pi \frac{M_{a_1}}{M_\rho} \ , \ \gamma= -\frac{F_\pi}{\sqrt{2}} \ .
\end{equation}
One can verify that this current satisfies PCAC.

At 1-loop level, multi-pion states contribute which render the Noether current more
involved. We follow Refs.~\cite{Weinberg:1996kr,Urban:2001ru} to infer a generalized
form of PCAC from the first Weinberg sum rule~\cite{Weinberg:1967kj},
\begin{equation}
\label{eq:pcac}
\int \frac{(-{\rm Im} \Pi_A^L)}{\pi s} ds = \frac{F_\pi^2}{2}  \ ,
\end{equation}
where $\Pi_A^L$ is the 4-D longitudinal axial current-current correlator.
It is defined by considering all possible $W$-$W$ transitions:
the direct $W$ selfenergy, $\Sigma_{WW}$, as well as $W$ couplings through
$a_1$ and/or $\pi$ propagators,
\begin{equation} \label{eq:imaL}
\begin{split}
\Pi_A^L &=  \Sigma_{WW} + \frac{\left(\beta + \Sigma_{Wa}\right)^2}{M_a^2 -
\Sigma_{aa}^L} + \frac{p^2}{p^2 - M_\pi^2 - \Sigma_{\pi}}
\\
&\times
\left( -\frac{F_\pi}{\sqrt{2}}+\Sigma_{W\pi} -\frac{\Sigma_{a\pi}}{M_a^2-\Sigma_{aa}^L}
\left(\beta+\Sigma_{Wa}\right) \right)^2 \ .
\end{split}
\end{equation}
The $\Sigma_{ij}$ are selfenergies with external legs $i,j$=$a_1,\pi,W$;
the pion selfenergy includes a direct contribution and one with an intermediate
$a_1$ propagator,
\begin{equation}
\Sigma_{\pi} \equiv \Sigma_{\pi\pi} + \frac{p^2 \Sigma_{a\pi}^2}{M_a^2 - \Sigma_{aa}^L} \ .
\end{equation}

One can now show that Eq.~(\ref{eq:pcac}) is satisfied when the
selfenergies obey the relations
\begin{equation}
\label{eq:longrel}
\begin{split}
\Sigma_{WW}& = \frac{\alpha^2}{\beta^2} \Sigma_{aa}^L, \ \,
\Sigma_{Wa}= -\frac{\alpha}{\beta} \Sigma_{aa}^L, \ \Sigma_{W\pi} = \frac{\alpha}{\beta} \Sigma_{a\pi},\\
\Sigma_{aa}^L&= \frac{\beta^2}{\gamma^2 p^2}\Sigma_{\pi\pi}, \
\Sigma_{a\pi}= -\frac{\beta}{\gamma p^2} \Sigma_{\pi\pi} \ .
\end{split}
\end{equation}
(note their scaling with the tree-level couplings).
The first three are a consequence of gauge symmetry, while the last two encode chiral
symmetry. Let us therefore focus on the latter: our strategy is to use them as criteria
to enforce PCAC.

\begin{figure}[!tb]
\centering
\includegraphics[width=.4\textwidth]{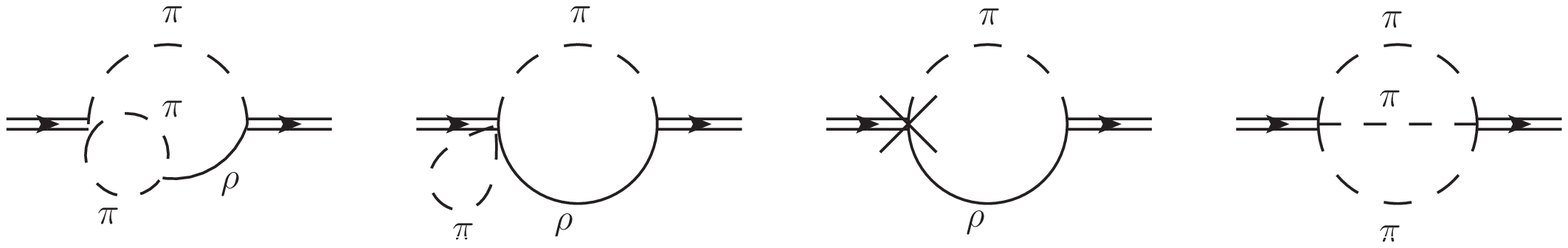}
\caption{Vertex correction diagrams to recover PCAC in the presence of a full $\rho$ propagator.
External legs are $W$, $a_1$ or $\pi$ to generate the selfenergy $\Sigma_{ij}$.
The first three diagrams show the corrections to the left-hand vertex while the complete set
includes the corrections to the right-hand vertex and to both simultaneously.
The $\rho$ propagator in each diagram is the fully dressed one.}
\label{fig:vertexdia}
\end{figure}

At 1-loop level, where the $\rho$-meson figures with a ``sharp" (bare) mass $M_\rho$ in the $\pi\rho$
loops, the longitudinal selfenergies can be factored into a universal loop integral, $\bar{\Sigma}$,
and vertex functions, $v_{ij}$, as
\begin{equation}
\Sigma_{aa}^L = v_{aa}(M_\rho^2) \bar{\Sigma} \ , \ \Sigma_{a\pi}= v_{a\pi}(M_\rho^2)\bar{\Sigma} \ , \
 \Sigma_{\pi\pi} = v_{\pi\pi}(M_\rho^2) \bar{\Sigma} \ ,
\label{eq:sigl-fac}
\end{equation}
satisfying Eqs.~(\ref{eq:longrel}).
When using the full $\rho$ propagator, $D_\rho$, the selfenergies still factorize, but
the vertex functions now depend on both $M_\rho$ and the off-shell $\rho$ mass $M$ (to be
integrated over); {\it e.g.}, $\Sigma_{aa}^L$ takes the form
\begin{equation} \label{eq:broad-rho-se}
\Sigma_{aa}^L = \int \frac{- {\rm Im} D_\rho (M^2)}{\pi}\ v_{aa}(M_\rho,M) \ \bar{\Sigma} \ dM^2 \ .
\end{equation}
Since the $v_{ij}$ generally depend differently on $M$ and $M_\rho$, Eq.~(\ref{eq:longrel}) is
violated. However, it turns out that the inclusion of the vertex corrections depicted in
Fig.~\ref{fig:vertexdia} can recover the structure of Eq.~(\ref{eq:sigl-fac}), {\it i.e.},
selfenergies factorized into the same $v_{ij}(M_\rho^2)$ as before and a different but
universal loop integral, $\bar{\Sigma}^*$ (which now includes an additional $M$ integration).
This can be achieved by casting the first three diagrams into a vertex correction
proportional to $\Sigma_\rho$, and then combining it with the ``bare" 3$\pi$ loop
(last diagram in Fig.~\ref{fig:vertexdia}) where two pions are required to be in a relative
P-wave. One also has to make a judicious choice of the 3$\pi$, $\rho3\pi$
and counterterm couplings in Fig.~\ref{fig:vertexdia}, differing from the Lagrangian value. This is not
unexpected since the use of the full $\rho$ propagator implies a {\em partial} resummation in
the first place.

The energy dependence of the $a_1 \rightarrow \rho\pi$ width with the
dressed $\rho$ propagator and associated vertex corrections resolves the problem
of the zero, {\it cf.}~solid line in Fig.~\ref{fig:axialwidth}. As an
extra benefit, finite strength develops below the previously sharp
$\rho\pi$ threshold.  However, the concern over the high-energy behavior
persists, which we address next.

On dimensional grounds, the two-derivative $a_1\rho\pi$ coupling and the $\rho$
propagator render the $a_1$ selfenergy to scale as $p^6$ at high energies.
Since the onset of this behavior is only slightly above the $a_1$ peak, the large
width suppresses spectral strength in the AV spectral function causes it
to fall off faster than the $\tau$-decay data. We interpret this as an
indication that the effective hadronic theory is breaking down, and additional
physics is needed. In fact, this is precisely what has been deduced from a
recent phenomenological analysis~\cite{Hohler:2012xd} of V and AV spectral
functions, where a chirally invariant continuum has been postulated, parameterized
following \cite{Shuryak:1993kg} as
\begin{equation}
\rho_{\rm cont} (s) = \frac{1}{8\pi^2}\left(1+\frac{\alpha_s}{\pi}\right)
\frac{s}{1+\exp[(E_0-\sqrt{s})/\delta]} \ .
\end{equation}
This carries significant strength into the $a_1$ peak
region. When adopting this continuum in connection with our microscopic model, we are
able to overcome the lacking-strength problem of MYM.

\begin{figure}[!tb]
  \centering
        \includegraphics[width=.45\textwidth]{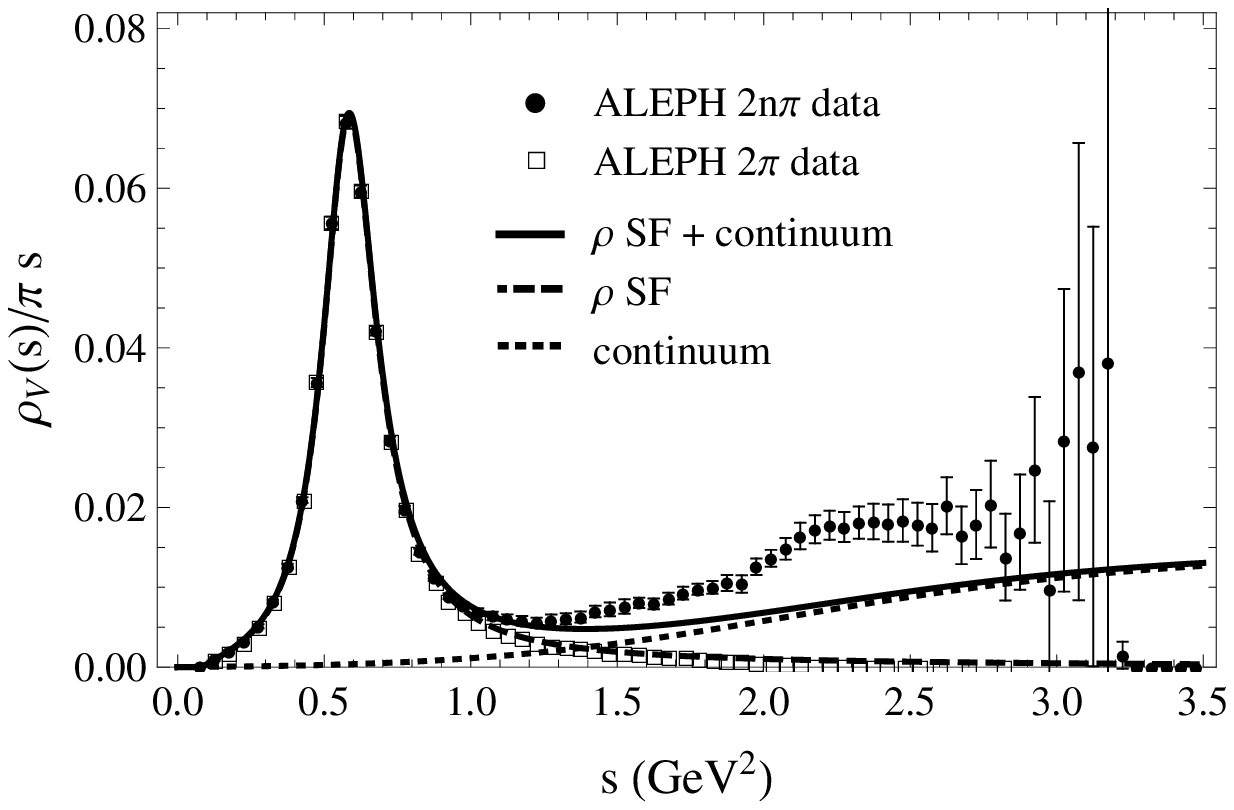}

\vspace{0.3cm}

        \includegraphics[width=.45\textwidth]{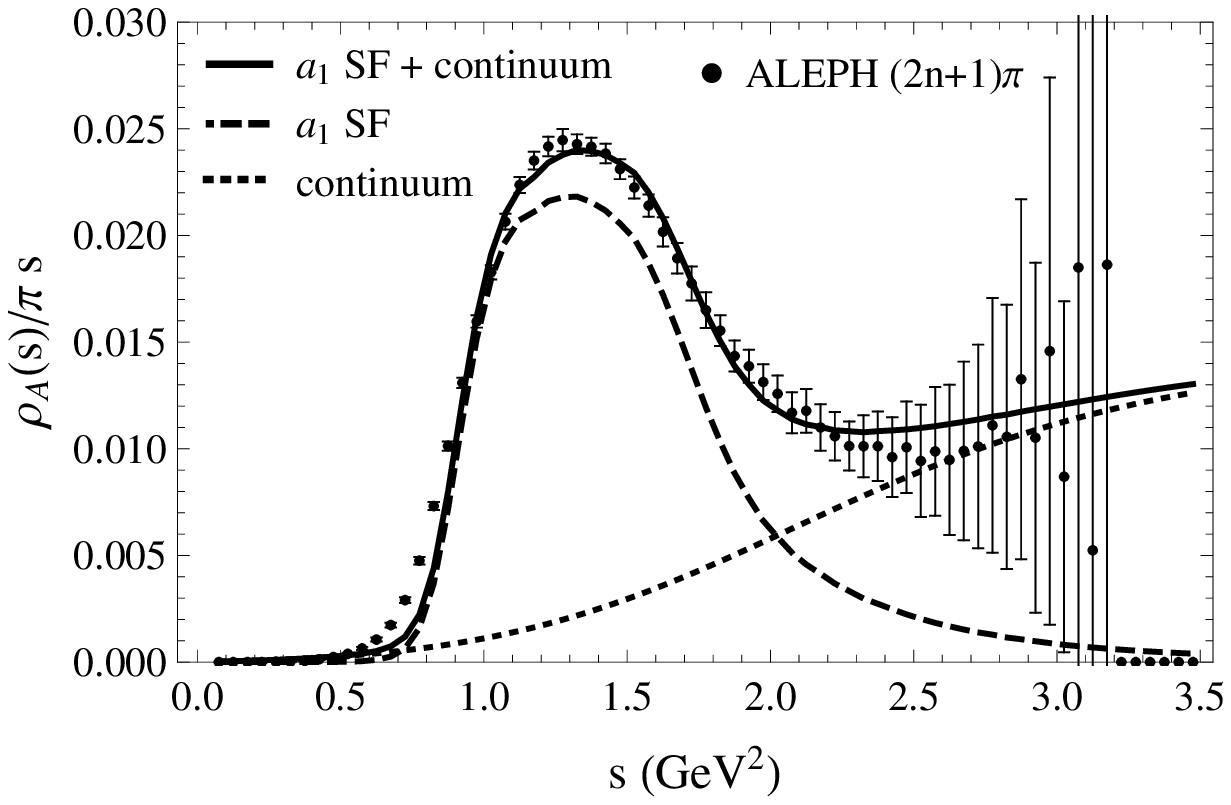}
\caption{Spectral functions in the vector (upper panel) and axial-vector
(lower panel) channels calculated from MYM (dashed lines), supplemented with
a chirally invariant continuum (dotted lines) and compared to $\tau$-decay
data~\cite{Barate:1998uf}.}
  \label{fig:sf}
\end{figure}
Our final fit to the $\tau$-decay data~\cite{Barate:1998uf}, shown in Fig.~\ref{fig:sf}, is
obtained with the parameters $g_{\rho\pi\pi}$=6.01, $g_{\rho\pi\pi}^{(3)}=$0.02\,GeV$^{-2}$,
$M_\rho$=0.86\,GeV, and  $M_{a_1}$=1.20\,GeV, with an additional six counter terms to
renormalize the theory. Note that $g_{\rho\pi\pi}^{(3)}$ is small as required for the
effective theory to have a meaningful derivative expansion.
The continuum parameters ($\alpha_s$=0.5, $E_0$=1.5\,GeV and $\delta$=0.2\,GeV)
have been slightly varied compared to Ref.~\cite{Hohler:2012xd} to optimize
the fit. Satisfactory agreement is found in both V and AV channels (excited states ($\rho'$
and $a_1'$) can be included to complete the fit~\cite{Hohler:2012xd}).
We have checked that data for the P-wave $\pi\pi$ phase shifts as well
as the pion electromagnetic formfactor
are reproduced well.
The calculated value for the radiative $a_1$ decay,
$\Gamma_{a_1 \rightarrow \rho\gamma}$=244\,keV is somewhat low compared to the
available data extracted from proton-nucleus collisions
(640$\pm$246\,keV)~\cite{Zielinski:1984au}.
Our D/S ratio of -0.131 (evaluated at the physical $a_1$ mass,
$\sqrt{s}$=1.23\,GeV) is significantly enhanced over the tree level result;
it approximately agrees with three of the four measurements reported by the
Particle Data Group~\cite{PDG:2013}.
We have numerically verified that PCAC is satisfied
through Eq.~(\ref{eq:pcac}) at the sub 0.1\% level, with the discrepancy
scaling with $M_\pi^2$, as expected.

We have conducted an analogous investigation within the linear realization of
chiral symmetry ($\sigma$ model). Using a sharp $\sigma$ meson, the fit to the
$\tau$ data preferred rather large masses ($m_\sigma$$\simeq$0.8\,GeV),
but did not reach the quality as in the non-linear version. Whether a broad
$\sigma$ meson can improve on this remains to be seen.

To summarize, we have set up a non-minimal MYM approach where, for the first
time, a resummed $\rho$ propagator has been implemented while maintaining
the chiral properties through suitably identified vertex corrections.
This enabled us to overcome previous problems in a realistic description of
the $\tau$-decay data in the axial-vector channel. The effective hadronic
theory decouples slightly above the $a_1$ peak, beyond which the recently
introduced idea of a universal perturbative continuum aids in maintaining
agreement with the experimental spectral functions.
We believe that our work helps to re-establish MYM as a viable description of
axial-/vector mesons in the chiral Lagrangian. This is particularly welcome
in the context of dilepton spectra in heavy-ion collisions, where the in-medium
vector channel (based on MYM) yields a quantitative description of available
data. The implementation of the here constructed axial-vector spectral
function into hadronic matter can thus serve as a quantitative framework to
establish the long-sought connection to chiral symmetry restoration.

\acknowledgments
We thank H. van Hees for discussions. This work is supported by the U.S.-NSF grant
no.~PHY-1306359 and by the A.-v.-Humboldt Foundation (Germany).



\begin{thebibliography}{99}
\bibitem{Gasser:1983yg}
  J.~Gasser and H.~Leutwyler,
  Annals Phys.\  {\bf 158}, 142 (1984).

\bibitem{Meissner:1987ge}
  U.~G.~Meissner,
  Phys.\ Rept.\  {\bf 161}, 213 (1988).

\bibitem{Gomm:1984at}
  H.~Gomm, O.~Kaymakcalan and J.~Schechter,
  Phys.\ Rev.\ D {\bf 30}, 2345 (1984).



\bibitem{Bando:1987br}
  M.~Bando, T.~Kugo and K.~Yamawaki,
  Phys.\ Rept.\  {\bf 164}, 217 (1988).



\bibitem{Harada:2003jx}
  M.~Harada and K.~Yamawaki,
  Phys.\ Rept.\  {\bf 381}, 1 (2003).


%
\bibitem{Song:1993ae}
  C.~Song,
  Phys.\ Rev.\ C {\bf 47}, 2861 (1993).

\bibitem{Ko:1994en}
  P.~Ko and S.~Rudaz,
  Phys.\ Rev.\ D {\bf 50}, 6877 (1994).

\bibitem{Barate:1998uf}
  R.~Barate {\it et al.}  [ALEPH Collaboration],
  Eur.\ Phys.\ J.\ C {\bf 4}, 409 (1998).

\bibitem{Ackerstaff:1998yj}
  K.~Ackerstaff {\it et al.}  [OPAL Collaboration],
  Eur.\ Phys.\ J.\ C {\bf 7}, 571 (1999).


\bibitem{Urban:2001ru}
  M.~Urban, M.~Buballa and J.~Wambach,
  Nucl.\ Phys.\ A {\bf 697}, 338 (2002).


\bibitem{Parganlija:2010fz}
  D.~Parganlija, F.~Giacosa and D.H.~Rischke,
  Phys.\ Rev.\ D {\bf 82}, 054024 (2010).



\bibitem{Wagner:2008gz}
  M.~Wagner and S.~Leupold,
  Phys.\ Rev.\ D {\bf 78}, 053001 (2008).


\bibitem{Hohler:2012xd}
  P.M.~Hohler and R.~Rapp,
  Nucl.\ Phys.\ A {\bf 892}, 58 (2012).

\bibitem{Weinberg:1996kr}
  S.~Weinberg,
  {\it The quantum theory of fields. Vol. 2: Modern applications},
  Cambridge Univ. Pr. (1996).


\bibitem{Weinberg:1967kj}
  S.~Weinberg,
  Phys.\ Rev.\ Lett.\  {\bf 18} (1967) 507.

\bibitem{Shuryak:1993kg}
  E.~V.~Shuryak,
  Rev.\ Mod.\ Phys.\  {\bf 65}, 1 (1993).







\bibitem{Zielinski:1984au}
  M.~Zielinski
{\it et al.},
  Phys.\ Rev.\ Lett.\  {\bf 52}, 1195 (1984).

\bibitem{PDG:2013}
J. Beringer {\it et al.} (Particle Data Group), Phys. Rev. D{\bf 86}, 010001 (2012).



\end{thebibliography}
\end{document}